# Resolving the size and charge of small particles: a predictive model of nanopore mechanics

Samuel Bearden[1], Tigran M. Abramyan[1], Dmitry Gil[1,2], Jessica Johnson[1], Anton Murashko[3], Sergei Makaev[3], David Mai[1,4], Alexander Baranchikov[5], Vladimir Ivanov[5], Vladimir Reukov[3*] and Guigen Zhang[1,6*]

[1] Department of Bioengineering, Clemson University, 301 Rhodes Hall, Clemson, SC 29634, [2]Massachusetts General Hospital, Harvard Medical School, Boston, MA 02114, USA, [3]University of Georgia, Athens, GA, 30602, USA [4]Department of Bioengineering, University of California, Berkeley, CA 94720, USA [5]Kurnakov Institute of General and Inorganic Chemistry, Russian Academy of Sciences, Leninskii pr. 31, 119991 Moscow, Russia, [6]F. Joseph Halcomb III, M.D. Department of Biomedical Engineering, University of Kentucky, 143 Graham Ave., Lexington, KY 40506, USA

**ABSTRACT**: The movement of small particles and molecules through membranes is widespread and has far-reaching implications. Consequently, the development of mathematical models is essential for understanding these processes on a micro level, leading to deeper insights. In this endeavour, we suggested a model based on a set of empirical equations to predict the transport of substances through a solid-state nanopore and the associated signals generated during their translocation. This model establishes analytical relationships between the ionic current and electrical double-layer potential observed during analyte translocation and their size, charge, and mobility in an electrolyte solution. This framework allows for rapid interpretation and prediction of the nanopore system's behaviour and provides a means for quantitatively determining the physical properties of molecular analytes. To illustrate the analytical capability of this model, ceria nanoparticles were investigated while undergoing oxidation or reduction within an original nanopore device. The results obtained were found to be in good agreement with predictions from physicochemical methods. This developed approach and model possess transferable utility to various porous materials, thereby expediting research efforts in membrane characterization and the advancement of nano- and ultrafiltration or electrodialysis technologies.

**GRAPHICAL ABSTRACT**

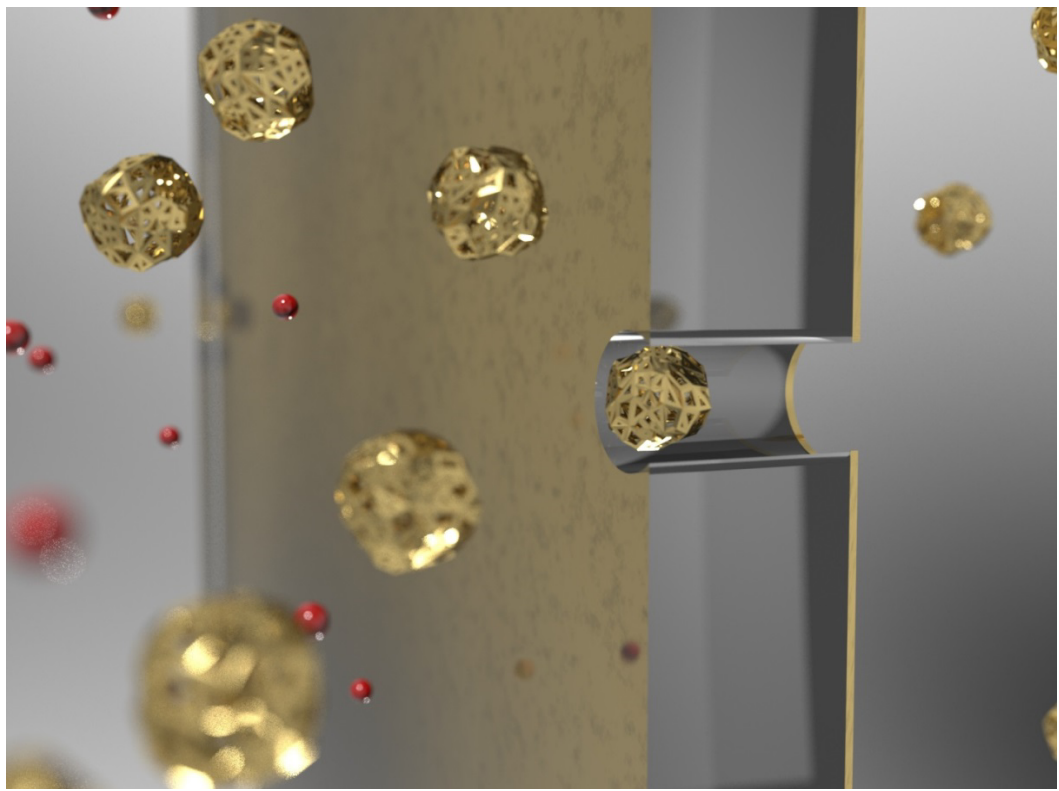

## 1. Introduction

Nanopore devices are complex systems with a wide application range, from nanofluidic valves and actuators [1-9] to filtration and separation devices [10-16] to high-resolution molecular sensors [17-20]. Furthermore, it includes techniques such as free-energy measuring [21] or automated dispersing of nanoparticles [22]. It is crucial to underline the importance of ion-selectivity characterisation via nanopore devices [23]. Special attention should be paid to the captivating and refined nanoporous membrane fabrication methods through the block copolymers' phase separation [24, 25].

Many mathematical models are developed based on different approaches [3, 17, 26-29] to increase the quality of devices and their applications. It would be remiss not to acknowledge that the finite-element method stands as one the most widely embraced techniques for such models, along with strategies based on molecular dynamics (MD) theory or combinations of both. For instance, Ethan Cao et al., in their research project Field [30], proposed the device and its mathematical description, striving to enhance the comprehension of cell membrane channels. Furthermore, Arjav Shah et al. put forward the universal approximation for conductance blockade in thin nanopore membranes based on a device for single particle transportation [30]. Notable-results were published by Sebastian Stock et al. about hydrogen densification in carbon nanopore confinement [31], which once again illustrates the breadth of research and its efficacy. Numerous models and experiments offer valuable insights into nanopore systems. However, few of them can interpreted as universal due to the vast diversity of tasks involved. Consequently, there is a need for the development of new models that emphasize universality and simplicity.

Thus, for instance, in our previous endeavour, we have meticulously crafted distinctive solid-state nanopores, featuring a splendid gold ring electrode adorning one aperture of a silicon nitride nanopore [32]. This ingenious contrivance grants us invaluable illumination into the underlying mechanisms of nanopore devices, elucidating the intricate interplay between the ionic current traversing the pore and the surface potential thereof [3]. In addition, if the surface potential of a metallic nanopore reaches equilibrium with the electrolyte solution, the potential appears to be mediated by the structure of the electrical double layer (EDL) [17, 33]. In particular, an analyte can pass the nanopore, which leads to disruption of the ionic equilibrium of the EDL structure and causes changes in both ionic current and EDL potential. Since ionic translocation through the EDL nanopore is hindered due to a gating effect, [3, 18] the analyte's dwell time often takes place in the range of milliseconds and is easily resolved by conventional electronics. The molecular signals in both ionic current and EDL potential are therefore affected by the size and charge of the transporting species, particularly when the size of the analyte species is much smaller than the nanopore diameter and wall interactions are not expected to dominate translocation characteristics. [17, 28]

An ability to link the measured ionic current and EDL potential signals of the physical and chemical structures of analyte species with a set of empirical equations would provide significant assistance in the design of nanofluidic devices and analysis of molecular analytes. With the concept firmly lodged in our minds, we embarked on the development process of a computer model. The model includes a set of empirical equations based on the thermodynamics principles governing the electric interactions and transport of ionic species in the nanopore. The model goal is to enable convenient and quantitative prediction of the properties of permeate by correlating the ionic current and EDL potential signals of analyte species to their size and charge, mediated by the strength of the electrolyte solution and the geometry of the nanopore.

## 2. Methods

### 2.1. Experimental system

The design and fabrication of a solid-state nanopore device, the experimental apparatus, and the signal extraction algorithms have been discussed in detail in our previous work [17]. Next, we'll briefly recap the key aspects of the previous work, nanopores with a diameter in the range of approximately 3 nm to 10 nm were created in a membrane consisting of a supporting layer of silicon nitride (50 nm thick) over a silicon substrate and a conducting metal layer of gold (5 nm thick). Nanopores were formed with electron-beam lithographic patterning and inductively coupled plasma (ICP) etching. Figure 1a shows a nanopore chip with a gold electrode overlaid on a suspended silicon nitride membrane, with an image of a nanopore formed through gold and silicon nitride layers. Nanopores may have conical, cylindrical, or hourglass geometries on a nanoscale, by the way, we chose to describe the geometry as a cylinder for modelling purposes. In addition, our solution was determined by the ICP etch fabrication technique. When placed in a flow cell with an aqueous electrolyte solution, the metal layer was electrically biased with a small constant electrical current ($I_{supp}$ = 37.4 ± 3.2 pA, Princeton Applied Research, Versastat MC, TN). Simultaneously, the ionic current through the nanopore (transmembrane current) was acquired (Molecular Devices, San Jose, CA) at 80,000 samples/s, which is sufficient to resolve translocations in this system. As depicted in Figure 1b, when an electrolyte solution (typically NaF) was driven through the nanopore, the corresponding ionic current was registered as the baseline-state ionic current ($I_{bs}$) and the EDL potential as the baseline-state EDL potential ($V_{bs}$) at the current source. When the electrolyte solution also contained analyte molecules as depicted in Figure 1c, the translocation of a single analyte molecule through the nanopore would cause spike signals to occur simultaneously in both the ionic current and EDL potential, and these signals were regarded as the perturbed-state ionic current and EDL potential, or $I_{ps}$ and $V_{ps}$, respectively. For quantifying the mobility of an analyte, its translocation time ($t_{ic}$) through the nanopore was determined by taking the full duration at half maximum (FDHM) of the recorded ionic current signal traces.

Firstly, the measurements were made for baseline-state $I_{bs}$ and $V_{bs}$ in a solution containing NaF, KCl, NaCl, LiF, or an equimolar mixture of NaF and KCl as electrolytes, each within an ionic strength range from $10^{-7}$ M to $10^{-1}$ M in logarithmic increments. Then measurements for perturbed-state $I_{ps}$ and $V_{ps}$ along with the associated translocation times ($t_{ic}$) were made for four previously characterized small molecule analytes, namely, citric acid, ascorbic acid, oxalic acid, and hydroquinone, at 10 nM in an electrolyte solution of NaF with the concentration of $10^{-7}$ M to $10^{-1}$ M in logarithmic increments.

For assessing the analytical capability of the developed model, two solutions containing colloidal nanocrystalline cerium dioxide ($CeO_2$) particles were also prepared and analysed. The ceria particles were synthesized according to the protocol adapted from the literature [34, 35]. Briefly, aqueous solutions of cerium (III) nitrate with different concentrations (0.1 – 0.5 M) were mixed with citric acid, and added dropwise to 3 M ammonia solution under constant stirring. The resulting purple suspension that corresponds to the formation of $(Ce^{+3}, Ce^{+4})O_y(OH)_z$ was kept at room temperature for 2 hours to facilitate oxidation and, thus, the formation of $CeO_2$ (ceria). The obtained ceria particles were rinsed several times with deionized water to remove an excess of ammonia and ammonium citrate.

**Table 1.** Size characteristics of the studied ceria samples.

| Name | Particle radius, DLS (nm) | Particle radius, TEM (nm) |
|---|---|---|
| "small" | 1.2 ± 0.2 | 1.7 ± 0.2 |
| "large" | 2.3 ± 0.8 | 3.5 ± 0.4 |

The sizes of the (Figure 1d, 1e, TEM, Hitachi H9500, Schaumburg, IL, USA). Table 1 lists the estimated sizes from both the DLS and TEM methods. These results show a significant difference in particle sizes between the two ceria samples as confirmed by two measurement techniques (DLS and TEM, both $p < 0.05$).

Before conducting nanopore experiments, the ceria stock solutions were diluted to a final concentration of 20 µM in 1 mM NaF at pH 2.1, and titrated with concentrated hydrochloric acid. Since redox properties of cerium oxide will change depending on the surrounding medium, the altered oxidized states of the nanoparticles were also evaluated after adding either microliter quantities of hydrogen peroxide solution (0.044 M $H_2O_2$) or ammonium hydroxide solution (0.044 M $NH_4OH$) to 10 ml aliquots of prepared nanoparticle solution.

z

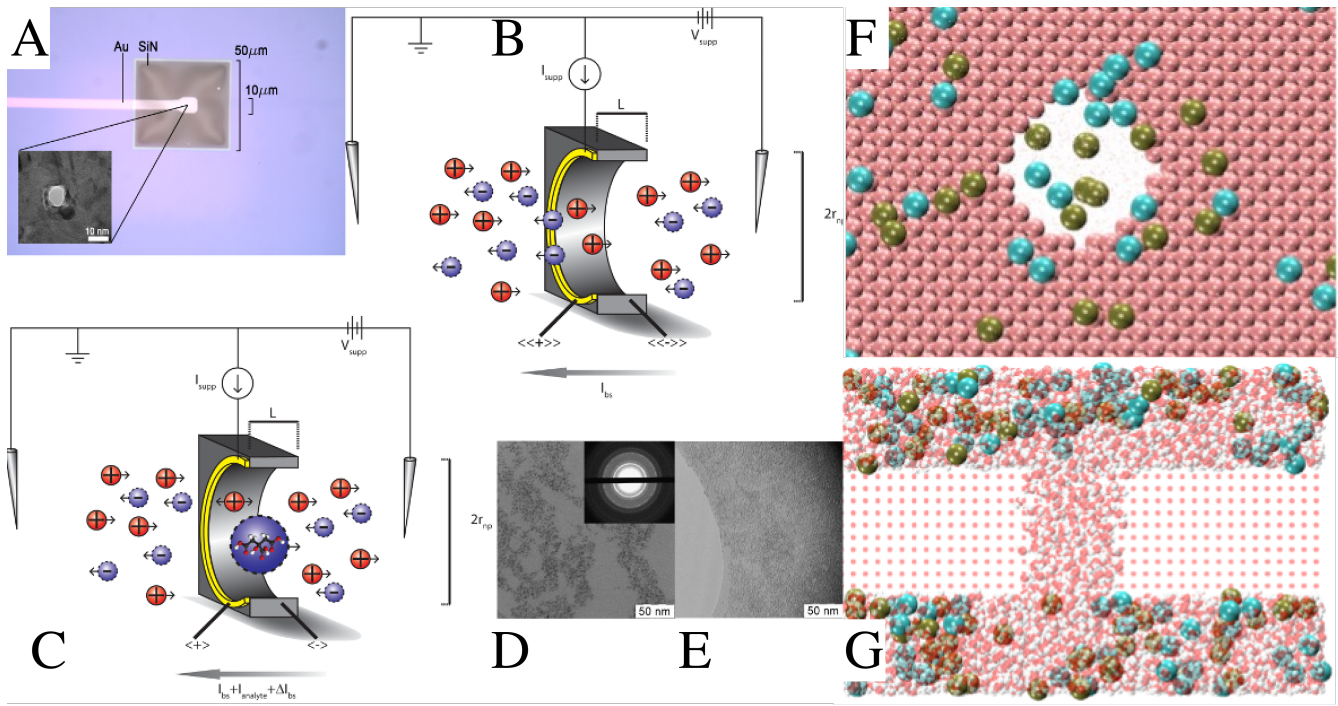

**Figure 1. a.** A nanopore chip contains a silicon nitride-suspended membrane overlaid with a gold electrode layer. The nanopore was formed through the gold and silicon nitride layers. **b.** A schematic diagram of an EDL nanopore in a baseline state in which a small biasing current ($I_{supp}$) is supplied to the gold ring electrode (represented in grey) and a constant cross-pore potential ($V_{supp}$) is applied across the nanopore. **c.** A schematic diagram of the nanopore in a perturbed state. Physical displacement and electrical interaction between the electrolyte and molecular analyte (dark grey) produce signals referred to as the perturbed-state EDL potential ($V_{ps}$) and ionic current ($I_{ps}$), respectively. **d.** TEM image of the "small" cerium oxide nanoparticles (1.7 nm radius). **e.** TEM image of the "large" cerium oxide nanoparticles (3.5 nm radius). The inset diffraction image confirms the phase composition of the sample. **f.** An MD model of a nanopore in an Au membrane is shown here from a top-down view along the nanopore axis. **g.** A side view orthogonal to that in 1f illustrates that ions were scattered throughout the equilibrated waterbox. No ions were placed inside the nanopore at the initial condition (t = 0 s).

**2.2. Molecular Dynamics study of the size partition effect in the baseline-state**

A Molecular Dynamics (MD) study was performed to explore the relationship between electrolyte ion size and baseline-state measurements (EDL potential and ionic current) as well as to visualize the process of ions entering the nanopore in the baseline state. In the MD study four situations were considered: NaF and KCl solutions in a nanopore device with its gold layer either partially charged or with no charge. For the partially charged cases, a valence charge of -0.2 was imposed on each gold atom.

The MD model of the gold surface was constructed using the Avogadro program [36]. A square-shaped unit cell in the x–y plane approximately 8 nm on each side (x and y directions) and 1.8 nm thick (z-direction) was generated. A channel with a radius of ~1 nm in the centre of the x–y plane along the z-axis was created by removing Au atoms to mimic experimental conditions (Figure 1f). CHARMM simulation program [37] (Chemistry at Harvard Macromolecular Mechanics, Harvard University, Boston, MA) was used for further model construction and simulations. The CHARMM22 protein force field [38] was used for the aqueous solution phase of the system (water and ions) and the metal force field [38] was used for the atoms of gold. A large water box was initially equilibrated at 1 bar pressure and 298 K temperature in an isothermal-isobaric ensemble with constant particle count (i.e., NPT ensemble) for 1.0 ns using the leapfrog integrator. The gold surface slab was then placed in the middle of the equilibrated water box. The water box was sliced to fit the size of the gold surface in the x–y plane, leaving a 1.7 nm solution layer on each side of the surface in the z-direction to contain water and ions. To simulate the experimental solution concentration of 100 mM, 81 molecules of salt ($Na^+$ or $K^+$ and $F^-$ or $Cl^-$, respectively) were then added to the water phase by randomly replacing water molecules with the atoms of the ions, avoiding placing ions in the gold channel (Figure 1g).

Three-dimensional periodic boundary conditions were applied to the MD simulation. The system was minimized using the steepest decent algorithm (first the gold surface keeping the solution phase constrained, then the solution phase locking the material surface). Then the gold atoms were constrained, and the rest of the system was equilibrated. MD production runs were performed in the canonical ensemble using the modified velocity-Verlet integrator [39] and a Nosé-Hoover thermostat. [40] Van der Waals (VdW) interactions were represented by 12-6 Lennard-Jones potential with a group-based force-switched cut-off, while the Coulombic interactions were represented using a group-based force-shift cut-off.

For both of the nonbonded interactions, the cut-off started at 0.8 nm and ended at 1.2 nm with a pair-list generation at 1.4 nm. SHAKE algorithm [41] was used to constrain the hydrogen bonds which enabled MD simulations with 2 fs timestep. For each system, simulation was performed for 2 ns and the frames were saved every 5 ps to monitor the entrance and behaviour of the ions in the nanopore.

## 3. Results and Discussion.
### 3.1. Case study: Predicting cerium oxide nanoparticle properties

To demonstrate the analytical capabilities of the model, the properties of ceria nanoparticles were characterized using our nanopore devices. For the ceria nanoparticles, we used a nanopore device with a radius of $r_{np}$ = 2.8 nm for the smaller nanoparticles and a nanopore device with a radius of $r_{np}$ = 4 nm for the larger nanoparticles. The measured ionic current and EDL potential signals for cerium oxide nanoparticles were statistically fit with the empirical relationships for ionic current signal ($I_{ps}$) and EDL potential signal ($V_{ps}$) to determine the radius ($r_{analyte}$) and charge ($z_{analyte}$) of the nanoparticles. The radius ($r_{analyte}$) and charge ($z_{analyte}$) are implicit in the charge density and charge velocity terms. The solution composition and nanopore geometry in the empirical equations were taken from the experimental conditions. Figure 2a shows the EDL potential signals obtained for the "small" (1-1.5 nm radius) cerium oxide nanoparticles and Figure 2b the corresponding ionic current signals, both are plotted alongside the error-minimized prediction. Figures 2c and 2d show the EDL potential and ionic current signals, respectively, for the "large" cerium oxide nanoparticles (2-3 nm radius) alongside the error-minimized predictions. For both the "small" and "large" nanoparticles, the EDL potential signals (Figure 2a, 2c) increase with the addition of $H_2O_2$ and quickly reach a maximum level, while the addition of $NH_4OH$ results in a decrease in signal magnitude to a minimum level. The ionic current signals for both types of nanoparticles (Figure 2b, 2d) exhibit an opposite trend with a decrease in current with the addition of $H_2O_2$ and an increase in current with the addition of $NH_4OH$.

Figure 2e shows the relationship between the predicted nanoparticle radius and the amount of $H_2O_2$ or $NH_4OH$ added for both the "small" and "large" nanoparticles. The radius predictions shown here are those producing the least residual error when comparing the predicted and measured ionic current and EDL potential signals, respectively. The radius of the "small" nanoparticles is predicted to be approximately 0.92 nm (1.84 nm diameter) in $NH_4OH$ and 1.27 nm (2.54 nm diameter) in $H_2O_2$. These values are consistent with the expected sizes of these nanoparticles (Table 1). The change in the radius of the ceria nanoparticles could be attributed to the formation of ceria-peroxo complexes on the particle surface. Alternatively, an increase in particle size can be the result of an increase in interatomic distances between Ce and O during the catalytic decomposition of hydrogen peroxide. [34, 42] In turn, the lengthening of the Ce-O bond can lead to the particle size increase. In contrast, the addition of $NH_4OH$ does not result in changes in interatomic distances, thus, the particle size remains lower than that in the presence of $H_2O_2$.

The predicted charge associated with the "small" nanoparticles (Figure 2f) is approximately 28 in $NH_4OH$ and 112 in $H_2O_2$. As with the predicted radii in Figure 2e, the charges shown in Figure 2f are those producing the least residual error when comparing the predicted and measured ionic current and EDL potential signals, respectively. The increase in the predicted charge is exactly 4 times for the "small" nanoparticles, as a result of oxidation of the surface cerium ions. The observed change in the surface could be attributed to the formation of various ceria-peroxo complexes on the particle surface upon exposure to the oxidizing agents. [43] Another point worth noting is that citrate ions that cover ceria nanoparticles are partially dissociated when exposed to the acidic $H_2O_2$ environment. This effect can also cause the change of surface charge measured in the present study. However, additional studies are required to determine the effect of citrate dissociation on surface charge alterations.

The predicted nanoparticle radius for the larger nanoparticles in a 4 nm nanopore varies between 0.95 nm and 1.59 nm (1.9 – 3.18 nm diameter, Figure 2e) and the predicted charge is between 9 and 60 (Figure 2f). In the evaluation of the larger nanoparticles, rather than considering the Stokes radius of the particle in the calculation of the perturbed-state volume effect ($\Delta n_{bs}$), the analyte radius ($r_{analyte}$) without a water layer is considered. Considering the Stokes radius of this larger nanoparticle in a larger nanopore results in dramatic under-prediction of the nanoparticle size (0.15 - 0.7 nm radius) compared with the size estimated from dynamic light scattering and TEM. Two effects could result in the under-prediction of the size of this nanoparticle when the Stokes radius is considered. Due to the increased radius of the nanoparticle, the water layer considered in the Stokes radius may be modified in such a way that it is no longer consistent with the assumptions of this model. Additionally, within the larger nanopore, the non-uniformity of the diffuse layer of the EDL will induce error in the calculation of the baseline-state charge density, where smaller nanopores should have a more uniform charge density ($n_{bs}$). It is worth noting that due to the assumptions in the model development in terms of small analytes and uniform charge density within the nanopore, among others, analytical accuracy may be enhanced for nanopores with radius up to 4 nm (diameter up to 8 nm). This enhancement at very small scales is unexpected but pleasant given the simplicity of the model and supports the accuracy and usefulness at very small length scales.

### 3.2. Empirical Equations Based on Experimental Observations

### 3.2.1. Empirical relationships for the baseline state

Based on the results of experiments, carried out by us, we notice the magnitude variation of the baseline-state EDL potential (Figure 2a) and ionic current (Figure 2d) depends on the concentration of the electrolyte solution in a similar decreasing trend. It appears that various electrolytes merely shift the curves along the vertical axis, and the polarity of the baseline states (both EDL potential and ionic current) is always opposite to the polarity of the smaller ion for each type of electrolyte. By exhaustively exploring the ratios of activities and diffusivity of the ions in these electrolyte solutions and comparing them with the experimentally measured potential curves, we noted that each of these variations (symbol plots in Figure 2a) represented a baseline state for the EDL potential associated with a specific electrolyte, which can be captured empirically by the following relationship:

$$V_{bs} = -sgn(z_{bs}) P_{bs} \frac{RT}{F} \frac{a_{reservoir}}{a_{nanopore}} \quad , (1)$$

with the nomenclature given in Table 2. The sign function (*sgn*) captures the polarity (+1 or -1) of the charged species. A negative sign is included in equation (1) since the surface potential ($V_{bs}$) has a polarity opposite of the majority ion within the nanopore. In equation (1), ionic activity (*a*) can be determined from concentration (*c*) and activity coefficient ($f_a$) through the relationship $a = cf_a$, where $f_a$ is determined using Debye's method as a function of the concentration, ionic radius, and valence charge of the ions in the solution at room temperature [44, 45]:

$$ln(f_a) = -0.849 z_i^2 \sqrt{\sum c_i z_i^2} \frac{1}{1 + 0.235 r \sqrt{\sum c_i z_i^2}} \quad , (2)$$

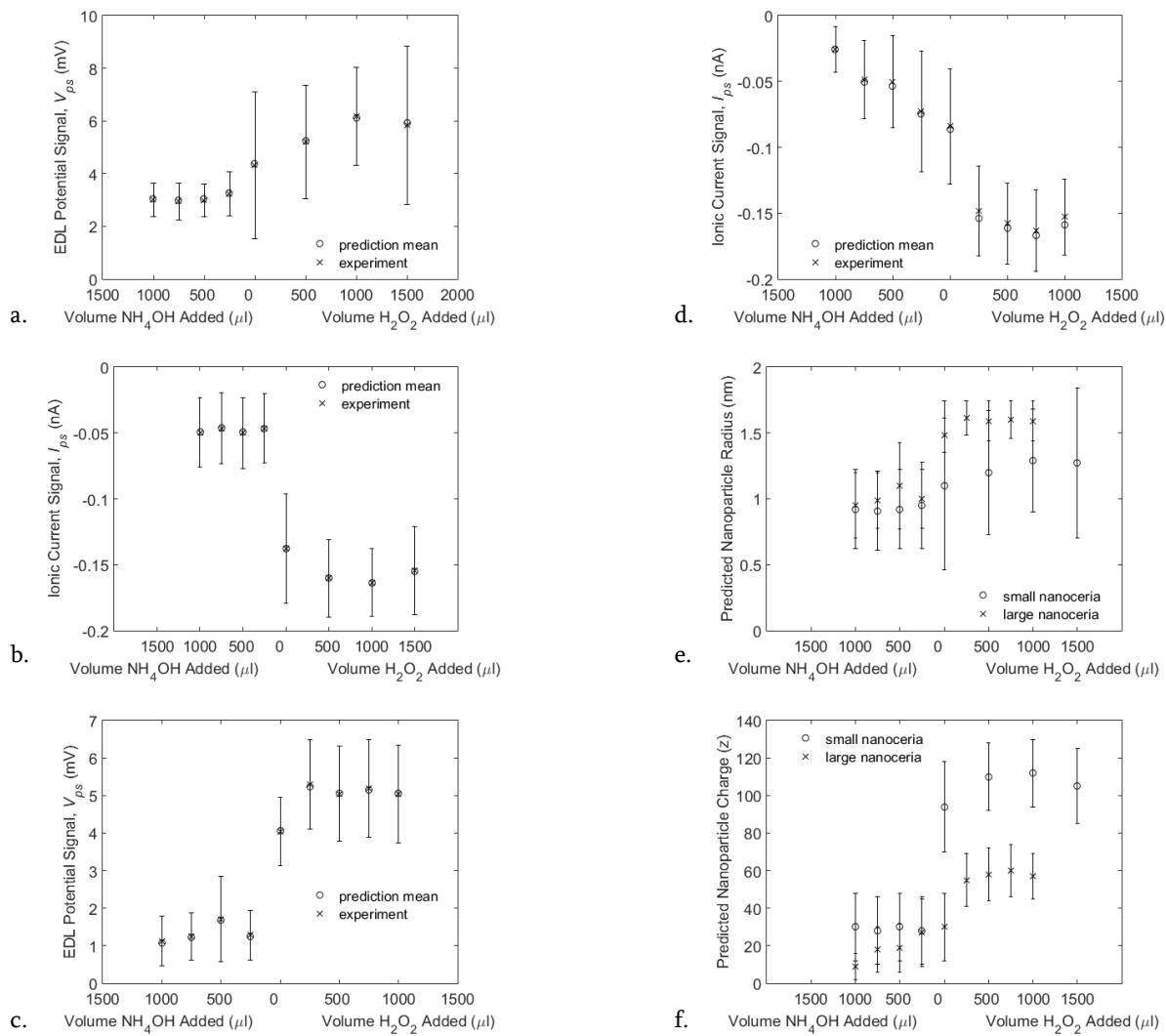

**Figure 2. a.** The (mean±SD) predicted and measured EDL potential signal for small cerium oxide nanoparticles from a nanopore with 2.3 nm radius in various concentrations of either ammonium hydroxide (NH₄OH) or hydrogen peroxide (H₂O₂). The predicted signal was fit to the experiment by adjusting the expected nanoparticle radius and charge. **b.** The ionic current signal for "small" nanoceria obtained simultaneously with the EDL potential signal and fitted by the model in the same way. **c.** The EDL potential signal for "large" nanoceria from a nanopore with 4 nm radius. **d.** The ionic current signal for large cerium oxide nanoparticles obtained simultaneously with the EDL potential signal and fitted by the model in the same way. **e.** The radius of the cerium oxide nanoparticles predicted from the model falls within the expected ranges of nanoparticle radii. Increased radius in the presence of hydrogen peroxide is consistent with expected changes to the crystal structure. **f.** The predicted charges of the cerium oxide nanoparticles become more positive in oxidizing solutions.

in which $r$ is the average ion radius, and $z_i$ and $c_i$ are the valence charge and concentration of species $I$, respectively. [43, 44] The ionic activity within the nanopore ($a_{nanopore}$) is calculated from the activities of the majority of ions within the nanopore (considering only the positive or negative ions), and the ionic activity within the reservoir ($a_{reservoir}$) accounts for all ions in the solution.

**Table 2.** Nomenclature.

| Symbol | Description | Unit |
| --- | --- | --- |
| $a_{nanopore}$ | Activity of solution within the nanopore | Mole L$^{-1}$ |
| $a_{reservoir}$ | Activity of solution within the reservoir | Mole L$^{-1}$ |
| $A$ | Cross sectional area of the nanopore | m$^2$ |
| $c_i$ | Concentration of species $i$ | Mole L$^{-1}$ |
| $C_{EDL}$ | EDL capacitance | F |
| $D$ | Diffusion coefficient | m$^2$ s$^{-1}$ |
| $D_{Kn}$ | Knudsen diffusion coefficient | m$^2$ s$^{-1}$ |
| $E$ | Electron charge | C |
| $\varepsilon$ | Permittivity | F m$^{-1}$ |
| $E$ | Driving electric field | V m$^{-1}$ |
| $f_a^i$ | Activity coefficient of species $i$ in solution | 1 |
| $F$ | Faraday's constant | C mole$^{-1}$ |
| $I_{analyte}$ | Ionic current due to the analyte | A |
| $I_{ps}$ | Ionic current signal | A |
| $I_{bs}$ | Baseline-state ionic current | A |
| $\Delta I_{bs}$ | Change to the current carried by the electrolyte solution | A |
| $k_B$ | Boltzmann's constant | m$^2$ kg s$^{-2}$ K$^{-1}$ |
| $L$ | Length of the nanopore | m |
| $L_{Au}$ | Length of the metal layer in the nanopore | m |
| $\mu$ | Mobility | m$^2$ V$^{-1}$ s$^{-1}$ |
| $\mu_{ion}$ | Mobility of the majority ion within the nanopore | m$^2$ V$^{-1}$ s$^{-1}$ |
| $n_{analyte}$ | Charge density of the analyte | C m$^{-3}$ |
| $n_{bs}$ | Baseline-state charge density within the nanopore | C m$^{-3}$ |
| $\Delta n_{bs}$ | Change to the charge density in the electrolyte solution | C m$^{-3}$ |
| $\Delta n_{bsE}$ | Change in charge density due to the charge of the analyte | C m$^{-3}$ |
| $\Delta n_{bsV}$ | Change in charge density due to the volume of the analyte | C m$^{-3}$ |
| $N_{Av}$ | Avogadro's number | Mole$^{-1}$ |
| $P_{bs}$ | Partition coefficient | 1 |
| $R$ | Ionic radius | m |
| $r_{analyte}$ | Ionic radius of analyte particle | m |
| $r_{Stokes}$ | Stokes radius of analyte particle | m |
| $R$ | Gas constant | J K$^{-1}$ mole$^{-1}$ |
| $T$ | Temperature | K |
| $t_{ic}$ | Translocation time measured in the ionic current signal | S |
| $V_{analyte}$ | Volume of the analyte | m$^3$ |
| $\overrightarrow{v_{analyte}}$ | Drift velocity of the analyte | m s$^{-1}$ |
| $V_{ps}$ | EDL potential signal | V |
| $V_{bs}$ | Baseline-state EDL potential | V |
| $\overrightarrow{v_{bs}}$ | Baseline-state drift velocity within the nanopore | m s$^{-1}$ |
| $\overrightarrow{\Delta v_{bs}}$ | Change to the drift velocity in the electrolyte solution | m s$^{-1}$ |
| $V_{supp}$ | Volt-clamp potential | V |
| $V_{total}$ | Volume inside the nanopore | m$^3$ |
| $z_{analyte}$ | Valence charge of the analyte | 1 |
| $z_i$ | Valence charge of species $i$ | 1 |
| $z_{bs}$ | Baseline-state majority ion valence | 1 |

It is expected that different electrolytes will give rise to different baseline-state EDL potentials, due to differences in the molecular weight and size of the ions involved. The partition coefficient ($P_{bs}$) used in equation (1) is to account for such molecular weight and size effects: $P_{bs} = \frac{D_{Kn,b}}{D_{Kn,a}} \frac{r_{np}}{L_C}$, where $D_{Kn,a}$ is the Knudsen diffusion coefficient $\left(D_{Kn} = \frac{2r_{np}}{3} \sqrt{\frac{8k_B T}{\pi M_w}}\right)$ of the majority ion within the nanopore, $D_{Kn,b}$ is the Knudsen diffusion coefficient of the minority ion within the nanopore, $r_{np}$ is the radius of the nanopore, and $L_C$ is the characteristic length scale of the system, which is taken as 1 nm in this study. Since the size of a molecule can be related to its molecular weight $r = \left(\frac{3}{4} \frac{M_w}{\pi N_{Av} \rho}\right)^{\frac{1}{3}}$, [46] the ratio of Knudsen diffusion coefficients is proportional to the ratio of ionic radii in an electrolyte pair: $\frac{D_{K,b}}{D_{K,a}} \propto \left(\frac{M_{w,a}}{M_{w,b}}\right)^{0.5} \propto \left(\frac{r_a}{r_b}\right)^{1.5}$. In quantifying the baseline-state EDL potential ($V_{bs}$), one could substitute the values for the activity of a solution ($a = cf_a$, via equation (2)) and for the partition coefficient ($P_{bs}$, determined by the ratio of the Knudsen diffusion coefficients and nanopore radius, $r_{np}$.) into equation (1) for a given electrolyte composed of ions at concentration $c_i$ with appropriate valences ($z_i$), along with molecular weights and other known physical constants (Table 3). The line plots in Figure 3a represent the various baseline-state EDL potentials predicted by equation (1) for the different electrolytes analysed. By plotting the measured baseline-state EDL potentials against the corresponding baseline-state ionic currents obtained from three nanopores of various sizes ($r_{np}$ = 1.7, 2.3, and 4.0 nm; Figure 3b), we observed a linear relationship. The linear relationship consists of three segments each made of data from a particular nanopore at various solution concentrations (between $10^{-7}$ M and $10^{-1}$ M inclusive of NaF). Through statistical regression ($R^2$ = 0.9735), we obtained an empirical expression for the linear relationship between the baseline-state ionic current ($I_{bs}$, in units of amperes) and the baseline-state EDL potential ($V_{bs}$, in units of volts) as:

$$I_{bs} = 10^{-8} V_{bs} - 4.8 * 10^{-10}$$

(1)

Figure 3c shows the baseline-state EDL potential and ionic current predicted from equations (1) and (3) for nanopores of radius 1.7, 2.3, and 4 nm in solutions of NaF between $10^{-7}$ and $10^{-1}$ nM inclusive. Note that Figure 3c closely resembles Figure 3b in all three segments of data from nanopores of three different sizes at various solution concentrations. The baseline-state EDL potential from equation (1) can be calculated from information on experimental conditions (electrolyte type, concentration, and nanopore radius) as discussed earlier in this section (**3.2.1**), and ionic current may be calculated with the potential as ($V_{bs}$) the only independent variable. The resulting values of predicted baseline states (potential versus current) are obtained from experimental conditions and equations (1) and (3) only. Extending this procedure to the different electrolyte solutions in Figure 3d shows the measured, as well as the predicted, baseline-state ionic current data for the various electrolyte solutions considered, varying with solution concentration over a 6 order-of-magnitude range. Note that measurements of the baseline-state ionic current (symbolic scatter plots in Figure 3d) were acquired simultaneously with those of the baseline-state EDL potentials shown in Figure 3a.

Similar to the potential (Figure 3a), the current (Figure 3d) is offset in magnitude by the type of electrolyte in the solution and de-creases in magnitude as the concentration increases. A good agreement can be seen between the measured and predicted current curves. Taken together, the measured baseline-state EDL potential and ionic current can be related, through equations (1) and (3), to the size of the nanopore and the type and concentration of the electrolyte solution.

Table 3. Experimental values.

| Symbol | Description | Value | Unit |
|---|---|---|---|
| $M_{w,Li}$ | Molecular weight of lithium ion | 6.94 | g/mole |
| $M_{w,F}$ | Molecular weight of fluoride ion | 18.99 | g/mole |
| $M_{w,Na}$ | Molecular weight of sodium ion | 22.99 | g/mole |
| $M_{w,Cl}$ | Molecular weight of chloride ion | 35.45 | g/mole |
| $M_{w,K}$ | Molecular weight of potassium ion | 39.1 | g/mole |
| $M_{w,OA}$ | Molecular weight of oxalic acid | 90.03 | g/mole |
| $M_{w,HQ}$ | Molecular weight of hydroquinone | 110.1 | g/mole |
| $M_{w,AA}$ | Molecular weight of ascorbic acid | 176.1 | g/mole |
| $M_{w,CA}$ | Molecular weight of citric acid | 192.1 | g/mole |
| $r_{Li}$ | Radius of lithium ion | 0.09 | nm |
| $r_F$ | Radius of fluoride ion | 0.119 | nm |
| $r_{Na}$ | Radius of sodium ion | 0.116 | nm |
| $r_{Cl}$ | Radius of chloride ion | 0.167 | nm |
| $r_K$ | Radius of potassium ion | 0.152 | nm |
| $r_{OA}$ | Radius of oxalic acid | 0.27 | nm |
| $r_{HQ}$ | Radius of hydroquinone | 0.32 | nm |
| $r_{AA}$ | Radius of ascorbic acid | 0.35 | nm |
| $r_{CA}$ | Radius of citric acid | 0.37 | nm |
| $z_{Li}$ | Valence of lithium ion | 1 | |
| $z_F$ | Valence of fluoride ion | -1 | |
| $z_{Na}$ | Valence of sodium ion | 1 | |
| $z_{Cl}$ | Valence of chloride ion | -1 | |
| $z_K$ | Valence of potassium ion | 1 | |
| $z_{OA}$ | Valence of oxalic acid | -1 | |
| $z_{HQ}$ | Valence of hydroquinone | -1 | |
| $z_{AA}$ | Valence of ascorbic acid | -2 | |
| $z_{CA}$ | Valence of citric acid | -3 | |

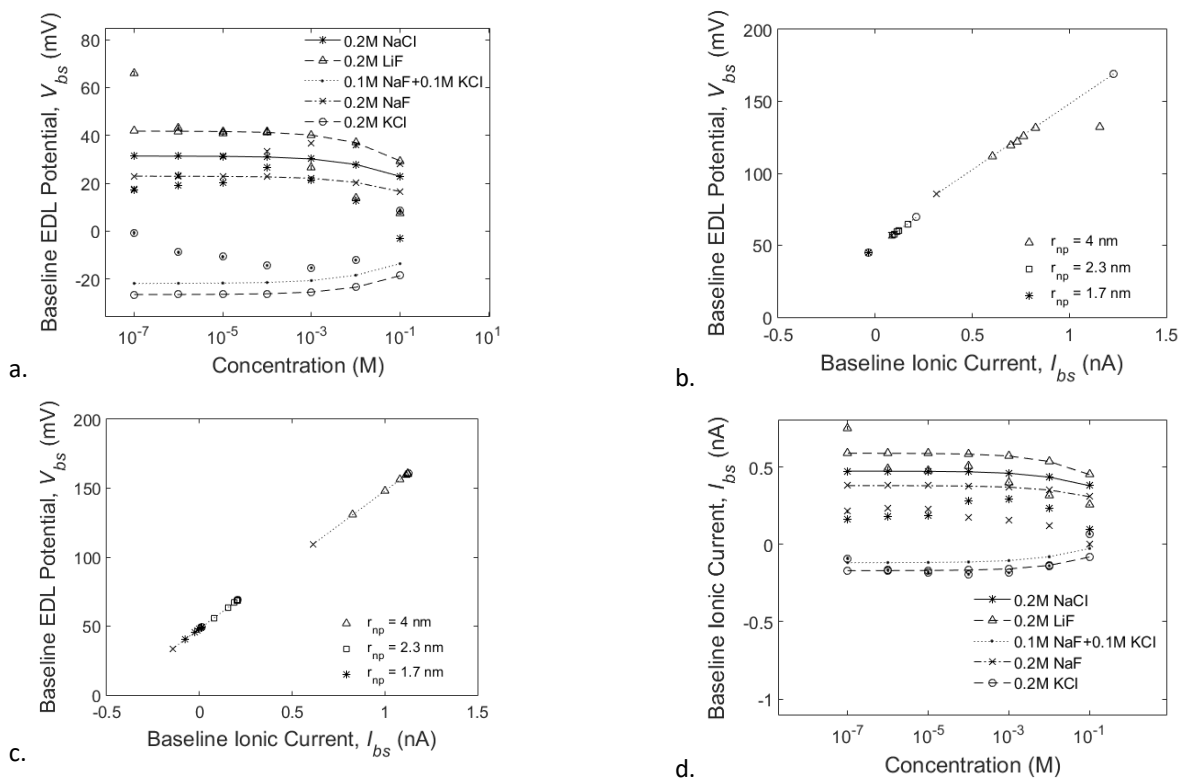

**Figure 3. a.** The baseline-state EDL potential is found to decrease in magnitude in a similar trend as the concentration of the electrolyte solution increases, regardless of the type of electrolytes used. The corresponding predicted EDL potentials (line curves) and the measured potentials (symbol curves) agree well with each other. **b.** Scatter plots of the baseline-state ionic current vs EDL potential obtained by pooling together all results for nanopores with different radii (1.7, 2.3 and 4 nm) and solution concentrations from $10^{-7}$ M to $10^{-1}$ M. Note that for each nanopore, the low concentration ($10^{-7}$M) is marked by an 'x' and the high concentration ($10^{-1}$ M) is marked with an 'o' to highlight the data segments belonging to different nanopores. **c.** Similar scatter plots as in 3b were predicted based on the empirical relationships (equations (1) and (3)). **d.** The baseline ionic current exhibits a similar behaviour as the baseline-state EDL potential in 3a.

### 3.2.2. Molecular dynamics results

The baseline-state EDL potential seems to result partly from a size-selection effect against larger ions in the nanopore. A trend was noted, indicating that the polarity of the baseline-state potential was opposite to that of the smaller ion in the electrolyte pair. In the MD model containing NaF in an uncharged nanopore, fluoride ($F^-$) enters the nanopore first, since there is a large size difference between the ions (Figure 4a). When a strong negative charge is applied to the nanopore in NaF solution, the positive sodium ($Na^+$) ions make up the charge within the nanopore due to electrostatic interaction (Figure 4b) which is dissimilar to the experimental measurement and the high positive ion density occurs because of the artificially fixed charge distribution in simulation. When considering KCl in an uncharged nanopore, both ions enter the nanopore at similar rates because the ions are very similar in size and there is no electrostatic selection (Figure 4c). As with the charged NaF model, when a charged nanopore is evaluated with KCl solution, the positive ion makes up the majority of the charge due to electrostatic interactions (Figure 4d) which is consistent with the expectation that the majority ion is a cation in KCl solution. From these simulations, it can be seen qualitatively that when ionic radii of the electrolyte are sufficiently different, the smaller ion enters the nanopore first, inducing the metallic nanopore surface to carry a potential of opposite polarity which continues charging until equilibrium is reached. When the ions are similar in size, there must be an initial electrostatic effect that selects for positive ion polarity before charging to the equilibrium potential, possibly due to the proximity of the SiN portion of the nanopore, which carries a negative polarity. However, to be consistent with experimental observation, the electrostatic selection effect must be considerably weaker than the size selection effect. These findings align with the patterns observed in the experimental data, where it's typically noted that the smaller ion exhibits polarity opposite to that of the baseline-state EDL potential. (Figure 4a). The MD results indicate that ion size plays a role in determining the polarity of the baseline-state potential and current, especially when there's a substantial disparity in size between positive and negative ions. However, when the sizes are comparable, additional electrical effects may contribute to ion separation, favouring the influx of positive ions into the nanopore. (Figure 4). Experimentally, the magnitude of the ionic valence charge was not a confounding factor since all electrolytes considered were monovalent pairs, indicating that any differences in baseline-state polarity are primarily due to the size effect.

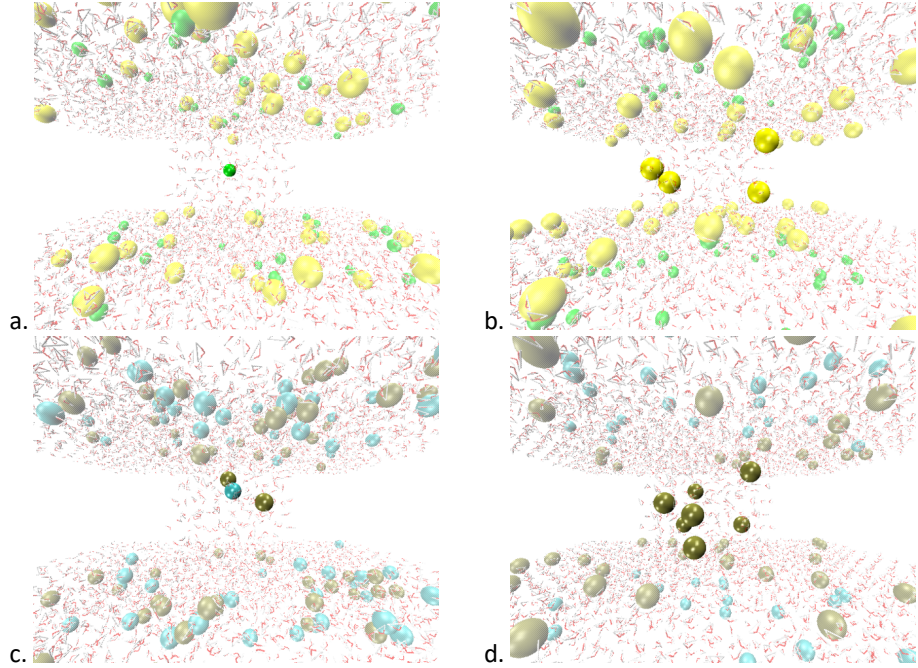

**Figure 4. a.** When the gold layer is uncharged, MD results predict that the smaller F$^-$ ions in the NaF solution will preferentially enter the nanopore first. Here, Na$^+$ ions are shown in yellow and F$^-$ ions are in green vdW representation, water is in transparent CPK representation. For clarity, the gold surface is not displayed, and the ions in the nanopore are shown in glossy vdW representation, whereas the ions outside the channel are shown in transparent vdW representation. **b.** When the gold layer is negatively charged, the larger positive ions, Na$^+$, in the NaF solution will enter the nanopore, overruling the preferential 'size effect'. **c.** Because the two ions in the KCl solution are of approximately similar size, the size selection has little effect in an uncharged nanopore. Here, K$^+$ ions are shown in tan and Cl$^-$ ions are in blue. **d.** In the negatively charged situation, the positive ions are selectively driven into the nanopore due to an electrostatic selection effect.

### 3.3. Further Predictions Based on the Empirical Relationships
### 3.3.1. The Baseline State

The baseline-state ionic current ($I_{bs}$) described empirically by equation (3) can also be expressed in terms of the velocity and charge density of the electrolyte solution moving through the cross-section of a nanopore (Figure 1b):

$$I_{bs} = sgn(z_{bs})\, n_{bs} A \vec{v_{bs}} \quad (4)$$

where $n_{bs}$ is the net charge density within the nanopore (in units of C/m$^3$), $A$ is the cross-section area of the nanopore and $\vec{v_{bs}}$ is the drift velocity of the net charge. In this expression, $\vec{v_{bs}}$ can be determined from the electric field across the nanopore and the mobility of the majority ion:

$$\vec{v_{bs}} = -sgn(z_{bs})\, \mu_{ion} E F \quad (5)$$

where $\mu_{ion}$ is the mobility of the majority ion and $E$ is the electric field which is determined as the baseline-state EDL potential over the length of the metal layer ($E = V_{bs}/L_{Au}$).

With equations (3) and (4), the charge density ($n_{bs}$) within the nanopore can be determined as:

$$n_{bs} = sgn(z_{bs}) \frac{V_{bs} - 0.048}{10^8 A \vec{v_{bs}}} \quad (6)$$

By substituting equations (5) and (6) into equation (4) we can describe the baseline-state ionic current in terms of the baseline-state EDL potential, valence, and velocity of the majority of ions in the solution.

### 3.3.2. The Perturbed-State

An analyte can translocate the nanopore, causing the production of a perturbed state (Figure 1c). The deviation from the baseline state can be prompted by disruption of the baseline-state charge density, baseline-state charge velocity, or direct contribution of the analyte. The charge density change in the electrolyte is caused by partial occlusion of the nanopore by the analyte molecule and compensatory charge accumulation due to electrostatic interaction with the charged analyte. The volumetric occlusion is known as the blockade effect and is commonly considered as the primary source of the ionic current signal in nanopores. We express the perturbation of the baseline-state charge density of the electrolyte within the nanopore as the sum of the changes in charge density caused by volumetric and electrical interactions between the analyte and electrolyte solution:

$$\Delta n_{bs} = \Delta n_{bsV} + \Delta n_{bsE} \quad (7)$$

The charge density of the electrolyte is altered through partial occlusion of the nanopore by the analyte molecule ($\Delta n_{bsV}$) and compensatory charge accumulation which is caused by electrostatic interaction with the charged analyte ($\Delta n_{bsE}$).

The change in charge density due to analyte volumetric occlusion can be determined by considering the amount of baseline-state charge that must be displaced:

$$\Delta n_{bsV} = -\frac{n_{bs} V_{analyte}}{V_{total}} \quad (8)$$

In this case, the volume of the analyte ($V_{analyte}$) covers a portion of the total nanopore volume, and the total charge within the nanopore is reduced by the amount of ionic charge that occupies the analyte volume in the baseline state. The analyte volume is calculated in this study as a sphere with the radius ($r_{Stokes}$) defined as the average radius of the analyte ($r_{analyte}$) with a water layer (the Stokes radius).

The change in charge density within the nanopore caused by electrostatic interaction between ions and the analyte can be expressed as:

$$\Delta n_{bsE} = -\frac{z_{analyte} e}{\pi r^2 L} \quad (9)$$

Equation (9) holds in both cases where ions are either attracted or repelled by the charged analyte.

The change in electrolyte drift velocity and ($\Delta \vec{v_{bs}}$) can then be determined using equation (6) with consideration of the perturbed-state charge density ($n_{ps} = n_{bs} + \Delta n_{bs}$) and drift velocity ($\vec{v_{ps}} = \vec{v_{bs}} + \Delta \vec{v_{bs}}$):

$$\Delta \vec{v_{bs}} = \frac{sgn(z_{bs})(V_{bs} - 0.048)}{10^8 A(n_{bs} + \Delta n_{bs})} - \vec{v_{bs}} \quad (10)$$

It should be noted that both the change in charge density ($\Delta n_{bs}$) and change in velocity ($\Delta \vec{v_{bs}}$) are dependent on the baseline-state conditions ($n_{bs}$ and $\vec{v_{bs}}$), meaning that the magnitude of the molecular signals is modulated by the baseline-state.

### 3.3.3. Direct contribution of the analyte to the perturbed state

In a perturbed state (Figure 1c), where a single analyte molecule passes through the nanopore, spike signals are experimentally observed for the ionic current and EDL potential. [17] The magnitudes of the spike signals measured from the respective baselines are expected to be related to the size and charge of the analyte. [17] The direct contribution of the analyte to molecular signals ($I_{ps}$ and $V_{ps}$) may be considered separately from the effect of the analyte on the baseline-state charge density and velocity, where we consider that the perturbed state ionic current is the sum of the change to the baseline-state current and the current contribution of the analyte ($I_{ps} = \Delta I_{bs} + I_{analyte}$). To demonstrate the procedure for calculating the direct impact of the analyte on the ionic current signal (Ips), we consider the analyte's ionic current signal as arising from both the charge density and drift velocity of the analyte within the nanopore. The ionic current resulting from the direct influence of the analyte is depicted in a format akin to that of the baseline-state ionic current. (equation (4)):

$$I_{analyte} = sgn(z_{analyte}) n_{analyte} \vec{v_{analyte}} A \quad (11)$$

The valence charge of the analyte ($z_{analyte}$) and nanopore cross-sectional area (A) can be assumed known. By considering the analyte as a single charged particle within the nanopore, the analyte charge density ($n_{analyte}$) is calculated as:

$$n_{analyte} = \frac{z_{analyte}e}{\pi r^2 L} \quad (12)$$

The drift velocity of the analytes is determined directly by dividing the total length of the nanopore ($L$ = 55 nm) by the translocation time measured as full duration at half maximum (FDHM) of the ionic current signal:

$$\overrightarrow{v_{analyte}} = -\frac{L}{t_{IC}} \quad (13)$$

Although the change in charge density due to the presence of an analyte ($n_{analyte}$) is relatively large and will contribute to the EDL potential signal, because the analyte drift velocity ($\overrightarrow{v_{analyte}}$) is minimal, the product of the two in equation (11) leads to negligible ionic current signal ($I_{analyte}$). As such, the direct contribution of these drifting analytes to the ionic current signal (equation (11)) is found to be several orders of magnitude smaller than the total ionic current signal observed in the experiment. Therefore, the ionic current ($I_{ps}$) is regarded as mainly generated from the change in the baseline-state ionic current ($\Delta I_{bs}$), but the contribution of analyte charges to the EDL potential signal should not be neglected.

### 3.3.4. Molecular signals in the perturbed state

In a perturbed state, the ionic current signal is dominated by the change in the electrolyte current ($\Delta I_{bs}$) and it can be expressed in terms of the charge density ($\Delta n_{bs}$) and velocity of the electrolyte ($\Delta \overrightarrow{v}_{bs}$) within the nanopore (Figure 1c) in a similar form to equation (4) as:

$$\Delta I_{bs} = sgn(z_{bs})\Delta n_{bs}\Delta\overrightarrow{v_{bs}}A \quad (14)$$

The perturbed-state EDL potential can be calculated from the capacitance of the EDL and changes to the total charge in the nanopore as:

$$V_{ps} = \frac{\pi r^2 L(\Delta n_{bs} + n_{analyte})}{C_{EDL}} \quad (15)$$

where $n_{analyte}$ is the charge density of the analyte within the nanopore, equation (12). The capacitance of the EDL ($C_{EDL}$) can be found as the derivative of baseline-state charge in the nanopore (baseline-state charge is the product of the nanopore volume and charge density, $V_{total}n_{bs}$) concerning the EDL potential (Figure 4a):

$$C_{EDL} = \pi r^2 L \frac{\delta n_{bs}}{\delta V_{bs}} \quad (16)$$

The predicted EDL potential signal ($V_{ps}$, Figure 4b) and ionic current signal ($I_{ps}$, Figure 4c) exhibit similar trends and magnitudes as the measured values. For the EDL potential signals, while differing to a certain extent in magnitude, they overall fall in between -1 mV and 1 mV and follow a similar trend in both the predicted and measured data, with hydroquinone (HQ) possessing the most positive signals, citric acid (CA) the most negative signals, and oxalic acid (OA) and ascorbic acid (AA) in between. For the ionic current signals, however, the agreement between the predicted and measured data in terms of magnitude and varying trend appears weak (Figure 4c): they are in a similar bulk range in magnitude but experimental measurements lack consistent trends. While the underlying reason for this has yet to be elucidated, this result suggests that nanopore sensors relying on ionic current signal ($I_{ps}$) may inherently possess a high level of variability in trans-pore current signals, thereby implying that this kind of current variation may have hindered the development of trans-pore current based nanopore sensors since the mid-1990s. [47]

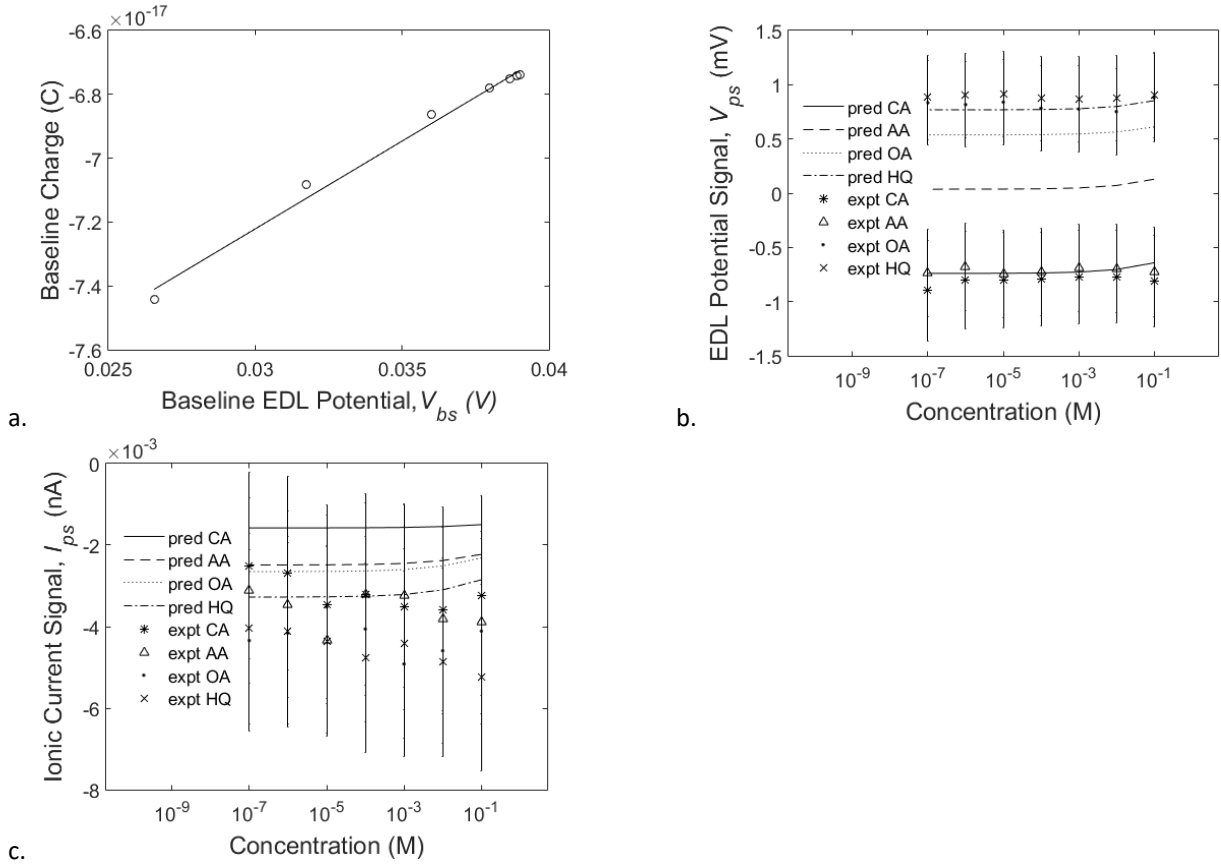

**Figure 4. a.** The baseline-state charge and EDL potential, predicted by the model, have a strong linear relationship. **b.** Predicted EDL potential signal compared with experimental measure (mean±SD). Range, signal order, and magnitude are similar between the predicted and experimental values. **c.** In comparison between the predicted and measured ionic current signals (mean±SD) for the four analytes used for validation, no clear trends are observed, but **the most** negative signals come from hydroquinone and the most positive (closest to 0 nA) from citric acid in both the predicted and experimental results.

### 3.3.5. Resolving the kinetic parameters of the analytes

Taking advantage of measurements of translocation time ($t_{IC}$), equation (5) can be rearranged along with the substitution of the driving electric field ($E = V_{bs}/L_{Au}$) to solve for the mobility of single molecules as a function of the translocation time and baseline-state EDL potential:

$$\mu = \frac{\Delta \overrightarrow{v_{bs}}}{EF} = \frac{L_{Au}L}{V_{bs}F t_{IC}} \quad (17)$$

With this mobility value, the corresponding diffusion coefficient as a function of the analyte valence can be determined by using the Stokes-Einstein relationship as follows:

$$D = \frac{\mu k_B T N_{Av}}{z_{analyte}} \quad (18)$$

This diffusivity can be quantified from information on known experimental conditions and parameters as well as the measured translocation time. As it is independent of the EDL potential signal ($V_{ps}$) and ionic current signal ($I_{ps}$), it may be used, alongside the ionic current and EDL potential signals, as an additional parameter for identifying unknown molecular analytes, which can be useful, for example, *in-situ* analysis in electrodialysis setups. Figure 5 shows the obtained values of the mobility and diffusivity for various molecular analytes. The mobility values (Figure 5a) and diffusion coefficient values (Figure 5b) of the analyte molecules translocating through a nanopore are orders of magnitude smaller than what is typically reported in unconfined space. [48-50] This phenomenon could be attributed to the fact that these analytes must move through a charge and size-selecting region inside the nanopore, thereby restricting certain freedom of movement. Though the separation of signals is more prominent in the diffusion coefficient than in the mobility, both sets of signals for different analytes remain distinct in a wide range of electrolyte concentrations.

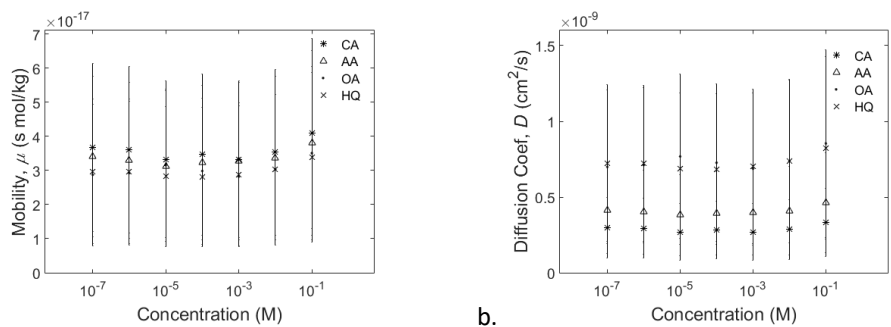

**Figure 5. a.** The (mean±SD) mobility of analyte ions vary with the type of species and concentration. **b.** The (mean±SD) diffusivity of analytes shows clear differences between species.

## 4. Conclusions

An analytical model, comprised of a series of empirical equations, has been crafted to encapsulate the operational mechanism of a nanopore fluidic device, drawing from experimental observations and the underlying principles of physics. These equations offer a direct linkage between the experimental parameters—such as the size and charge of analyte molecules and the electrolyte's solution strength — and the recorded signals of ionic current and EDL potential of the analyte molecules. Additionally, the mobility and diffusivity of analyte molecules within a nanopore can be precisely quantified, serving as supplementary parameters for characterising the molecular analytes.

In showcasing the functionality and efficacy of our model, we have successfully provided quantitative predictions for the size and charge of ceria nanoparticles susceptible to redox reactions in both acidic and basic environments. It is pertinent to highlight that our model, in contrast to many computational models, does not necessitate intricate and resource-intensive computations to facilitate its implementation—a notable enhancement. The interconnected equations governing this model hold considerable significance in illuminating the intricate behaviours of nanopore devices, such as membranes or selective sensors, thus fostering broader adoption of nanopore technology in propelling advancements in biomedical engineering and sciences.
**CORRESPONDING AUTHOR**

\* Corresponding Authors: Guigen Zhang, guigen.bme@uky.edu, (859) 323-7217, Vladimir Reukov, reukov@uga.edu

**AUTHOR CONTRIBUTIONS**

SB designed and performed experiments, developed models, analysed data and wrote the manuscript; TMA designed, performed and analysed the MD simulations. JJ, DM, and DG designed, performed and analysed experiments; DG, AB, VI, and VR prepared and characterized nanoparticles and edited the manuscript; AM and SM prepared the manuscript's formatting and edited the manuscript; VR and GZ supervised analysis and edited the manuscript.

**FUNDING SOURCES**

This work was completed with support from the Institute for Biological Interfaces of Engineering at Clemson University, Clemson University Department of Bioengineering, Clemson Computing and Information Technology, the Cyberinfrastructure Technology Integration group at Clemson University, and the Georgia Institute of Technology Institute for Electronics and Nanotechnology. The synthesis and characterization of ceria nanoparticles was supported by the Russian Science Foundation (grant 19-13-00416). This work is partially supported by funds from the Bill and Melinda Gates Foundation, an anonymous Venture group, and the National Science Foundation (NSF) Research Experience for Undergraduates award # EEC-1262991.

**Competing interests**

Samuel Bearden and Guigen Zhang are inventors on US Patent Application US20180202969A1.

**ACKNOWLEDGMENT**

**We would like to acknowledge Clemson University Palmetto cluster and its staff for support.**